\documentclass[nonacm,sigconf]{acmart}
\usepackage[utf8]{inputenc}

\usepackage{amsmath}
\usepackage{listings}
\usepackage{color}

\usepackage[inline]{enumitem}

\usepackage{todonotes}
\usepackage{bussproofs}
\usepackage{soul}
\usepackage{parcolumns}
\usepackage{color,soul}

\lstdefinestyle{customjava}
{language=java,
keywordstyle=\color{blue},
basicstyle=\ttfamily\small,
mathescape=true,
escapechar=|,
commentstyle=\color{dkgreen},   
frame=single
}

\definecolor{dkgreen}{rgb}{0,0.6,0}
\definecolor{gray}{rgb}{0.5,0.5,0.5}
\definecolor{mauve}{rgb}{0.58,0,0.82}

\title{SAT-Based Extraction of Behavioural Models for Java Libraries
  with Collections}

 \author{Larisa Safina}
 \email{larisa.safina@inria.fr}
 \affiliation{%
   \institution{Univ. Lille, Inria, CNRS, Centrale Lille, UMR 9189 CRIStAL}
   \country{F-59000 Lille, France}
 }

 \author{Simon Bliudze}
 \email{simon.bliudze@inria.fr}
 \affiliation{%
   \institution{Univ. Lille, Inria, CNRS, Centrale Lille, UMR 9189 CRIStAL}
   \country{F-59000 Lille, France}
 }

\setcopyright{acmcopyright}
\copyrightyear{2022}
\acmYear{2022}

\newcommand{\secn}[1]{Section~\ref{secn:#1}}
\newcommand{\fig}[1]{Figure~\ref{fig:#1}}
\newcommand{\figs}[2]{Figures~\ref{fig:#1} and \ref{fig:#2}}
\newcommand{\eq}[1]{(\ref{eq:#1})}
\newcommand{\hyp}[1]{(\ref{hyp:#1})}

\newcommand{\mdash}[1][]{{#1}---{#1}}
\newcommand{\eg}[1][\@ ]{e.g.{#1}}
\newcommand{\Eg}[1][\@ ]{E.g.{#1}}
\newcommand{\ie}[1][\@ ]{i.e.{#1}}
\newcommand{\etc}[1][\@ ]{etc.{#1}}
\newcommand{\resp}[1][\@ ]{resp.{#1}}
\newcommand{\cf}[1][\@ ]{cf.{#1}}

\newcommand{\sB}{\ensuremath{\mathbb{B}}}

\newcommand{\false}{\ensuremath{\mathit{false}}}

\newcommand{\bydef}{\ensuremath{\mathbin{\stackrel{\scriptscriptstyle\mathit{def}}{=}}}}

\newcommand{\setdef}[2]{\ensuremath{\{#1\,|\,#2\}}}
\newcommand{\bsetdef}[2]{\ensuremath{\bigl\{#1\,\bigr.\bigl|\,#2\bigr\}}}

\newcommand{\sgoesto}[2][]{\ensuremath{\xrightarrow[#1]{#2}}}
\newcommand{\goesto}[1]{\sgoesto{#1}}

\newcommand{\condition}{\ensuremath{\mathit{cnd}}}

\newcommand{\equal}[2]{\ensuremath{\mathit{eq}_{#1,#2}}}

\newcommand{\contains}[2]{\ensuremath{\mathit{cnt}_{#1,#2}}}
\newcommand{\empt}[1]{\ensuremath{\mathit{empty}_{#1}}}
\newcommand{\exception}{\ensuremath{\mathit{e}}}
\newcommand{\axiom}[1]{\ensuremath{\mathit{ax}_{\ref{ax:#1}}}}
\newcommand{\axioms}{\ensuremath{\mathit{axioms}}}

\newcommand{\api}{\ensuremath{\mathit{A}}}

\newcommand{\Collections}[1][{}]{\ensuremath{C_{#1}}}
\newcommand{\Values}[1][{}]{\ensuremath{V_{#1}}}
\newcommand{\Predicates}[1][{}]{\ensuremath{P_{#1}}}
\newcommand{\Context}[1][{}]{\ensuremath{(\Collections[#1], \Values[#1])}}
\newcommand{\State}[1][{}]{\ensuremath{\sigma_{#1}}}
\newcommand{\States}{\ensuremath{\Sigma}}

\newcommand{\code}[1]{\ensuremath{\mathtt{#1}}}
\newcommand{\method}{\code{meth}}

\newcommand{\encoding}[2]{\ensuremath{\mathit{enc}^{#2}(\code{#1})}}
\newcommand{\opformula}[1]{\ensuremath{f_{#1}}}
\newcommand{\opset}[1]{\ensuremath{\Predicates[#1]}}

\newcommand{\free}[1]{\ensuremath{{#1}_{\mathit{free}}}}

\newcommand{\last}[1]{\ensuremath{\code{#1}{\downarrow}}}

\DeclareMathOperator{\dom}{dom}
\newcommand{\domain}[1]{\ensuremath{\dom(#1)}}

\newcommand{\refines}[2]{\ensuremath{{#1} \prec {#2}}}

\newcommand{\rangestackii}[2]{%
  \renewcommand{\arraystretch}{0}%
  \begin{array}{@{}c@{}}\scriptstyle #1\\\scriptstyle #2\end{array}%
}
\newcommand{\rangestackiii}[3]{%
  \renewcommand{\arraystretch}{0}%
  \begin{array}{@{}c@{}}\scriptstyle #1\\\scriptstyle #2\\\scriptstyle #3\end{array}%
}

\newenvironment{textlist}{\begin{enumerate*}[label=(\arabic*)]}{\end{enumerate*}}

\newcommand\blfootnote[1]{%
  \begingroup
  \renewcommand\thefootnote{}\footnote{#1}%
  \addtocounter{footnote}{-1}%
  \endgroup
}


\begin{document}

\begin{abstract}
  Behavioural models are a valuable tool for software verification, testing, monitoring, publishing etc.
  However, they are rarely provided by the software developers and have to be extracted either from the source or from the compiled code.
  In the context of Java programs, a number of approaches exist for building behavioural models.
  Most of these approaches rely on the analysis of the compiled bytecode.
  Instead, we are looking to extract behavioural models\mdash in the form of Finite State Machines (FSMs)\mdash from the Java source code to ensure that the obtained FSMs can be easily understood by the software developers and, if necessary, updated or integrated into the original source code, \eg in the form of annotations.
  Modern software systems are huge, rely on external libraries and interact with their environment. Hence, extracting useful behavioural models requires abstraction.
  In this paper, we present an initial approach to this problem by focusing on the extraction of FSMs modelling library APIs.  We focus on the analysis of Java code involving the use of collections.
  To this end, we encode the operational semantics of collection operations using patterns of Boolean predicates.
  These patterns are instantiated based on the analysis of the source code of API implementation methods to form an encoding of the possible FSM transitions.
  A SAT solver is then used to determine the enabledness conditions (guards) of these transitions.
\end{abstract}

\maketitle


\section{Introduction}
\label{secn:intro}
\blfootnote{L. Safina and S. Bliudze were partially supported by ANR Investissments d’avenir (ANR-16- IDEX-0004 ULNE) and project NARCO (ANR-21-CE48-0011). P. van den Bos, M. Huisman and R. Rubbens were supported by the NWO VICI 639.023.710 Mercedes project.}

Behavioral models are a valuable tool for supporting the process of building the correct software, that could be beneficial during all phases of the development lifecycle: design and implementation~\cite{montesi15, voinea20, Bravetti19}, testing and verification~\cite{10.1145/332740.332741, burlo20}, coordination and monitoring~\cite{GabbrielliGLM19, Mavridou19}, deployment~\cite{Bravetti2020AFA} \etc

One of the particular problem behavioral models could help to address is maintaining the API correctness. APIs provide a great way of scoping, encapsulating and presenting library functions~\cite{bloch06}. Reusing components with the help of APIs helps to reduce the cost of software systems development and to increase their quality~\cite{Espinha2015WebAG, brito17, brito18}. However, API development is not free of problems. When designed badly, API offers inefficient or even incorrect code requiring writing additional code to deal with it, which leads directly to the increased development cost~\cite{Henning07}.  When documented badly and without clear explanation of  the correct sequence of method usage, API can demonstrate exceptional behaviours~\cite{ancona18}. External APIs, being a black box, can contain error-prone dependencies and make systems prone to breaking changes that have to be addressed in the client code. A new version of an external interface evolving independently can break backward compatibility and may cause runtime failures of client systems. According to studies, breaking changes are being present in the significant amount of all new releases (70\%~\cite{RAEMAEKERS2017140}, 75\%~\cite{ dietrich14}, and 80\%~\cite{dig06}). One can argue that building a dynamic model of libraries to help the developer understand the behaviour of methods operating on collections contradicts the fundamental principle of information hiding: a developer relying on knowledge of the API implementation would produce code even more prone to the breaking changes we want to address. We believe, however, that if a signature of an API method provides a developer with the insufficient information, she will need to make her assumptions in any case whether they are supported by the extracted model or not. But comparing the two models of two different releases can show a developer if any breaking changes were introduced.

There exist various approaches based on behavioral models to deal with such problems possessing different levels of formality (from graphical representation~\cite{Cook2017} to employing behavioral types~\cite{DBLP:journals/ftpl/AnconaBB0CDGGGH16}) and ways of being produced and presented. In general, these models are expected to be expressive, rigourous, intuitive, well-structured and possess the adequate level of abstraction (be sufficiently detailed and precise) \cite{harel97}. Such models could complement the code~\cite{Rumbaugh2004}, be extracted from it~\cite{dig06, 10.1145/2384616.2384623}, from its binaries~\cite{RAEMAEKERS2017140}, from some intermediate representation or from other models~\cite{luis17, dardha2017mungo}, from the execution traces~\cite{lorenzoli18} or be created to synthesize the code from them~\cite{harel97}. Our approaches follows the idea presented in \cite{henzinger05, Godoy21, 10.1145/2384616.2384623}. Henzinger et al. have introduced the idea of safe and permissive interfaces (which allow all possible sequences of method calls not violating interface's internal invariants and prohibits all others) and have presented an algorithm for authomatically synthesizing them as a graph with nodes as predicates over interface's internal states and labels as library calls. Godoy et al. were exploring search-based test generation techniques to exploit the abstraction of interfaces based on their enabledness (permitted sequennce of method calls) in order to find possible failures. Par\'izek et al. were working on verification of java programs with a particular focus on collections. Their approach is based on using a predicate language and the idea of weakest preconditions to model the states and possible method calls of a collection interface and capture possible changes. The novelty of our approach concerning presented above lies in the presence and absence of guards that help to break the information into the one known (in this case this information goes to states) and the one unknown (which goes to guards and resolves dynamically). Another consequence of our approach is that we can also achieve different levels of abstraction by adding more or fewer predicates and axioms.

We are interested in building behavioral models that have a tight connection with API implementation source code to provide feedback to developers on the behavior of the program. 
In this paper, we propose an approach to semi-automatic extraction of behavioral models for APIs written in Java in the form of Finite State Machines (FSMs), focusing on the code operating on collections. 

The paper is structured as follows. In \secn{example}, we introduce a simple running example.
In \secn{approach}, we present our approach to model extraction.
In \secn{discussion}, we discuss our assumptions and future work directions.  \secn{conclusion} concludes the paper.


\section{Running Example}
\label{secn:example}

Let's imagine the API created for the users of a pharmaceutical company. In particular it allows users to create reservations for the set of medications and to manipulate this collections: add or remove elements. Each medication can have two identifiers where the second is the one provided by a third-party vendor. When a detail to be added or removed both ids are checked (see the code snippet bellow)

\begin{figure}
\begin{lstlisting}[style=customjava]
  class ExampleImpl implements ExampleAPI {
    private Set<String> idSet;

    public ExampleImpl() {
1:    idSet = new HashSet<>();
    }
    
    public void add(String id){
1:      idSet.add(id);         // See |{\color{dkgreen}\secn{encoding}}| for the line
    }                          // numbers in the left margin

    public void removeId(String idMain, String idOpt) {
1:    idSet.remove(idMain);
      if (idMain != idOpt) {
2:      idSet.remove(idOpt);
      } else {
3:    }
    }
  }
\end{lstlisting}
  \caption{Running example source code }
  \label{fig:example}
\end{figure}

\begin{figure}
 \centering
 \includegraphics[width=0.35\textwidth]{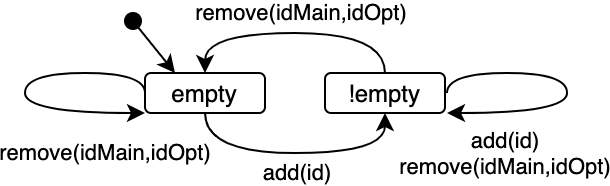}
 \caption{Behavioural model for the \code{HashSet} implementation}
 \label{fig:v1}
\end{figure}

During the last code review it was decided to migrate idSet collection from the \lstinline|HashSet| to the \lstinline|TreeSet| to present the order of the elements. However, this led to the unexpected exception, when some user tried to remove a medicine having only one id (\lstinline|remove("LK32EJ2", null)|). This exception is not visible at the compilation but can be catched by the static code analysers, and it is clearly visible if we compare the behavioral models generated for the two versions of the code (see \figs{v1}{v2}).

\begin{figure}
 \centering
 \includegraphics[width=0.40\textwidth]{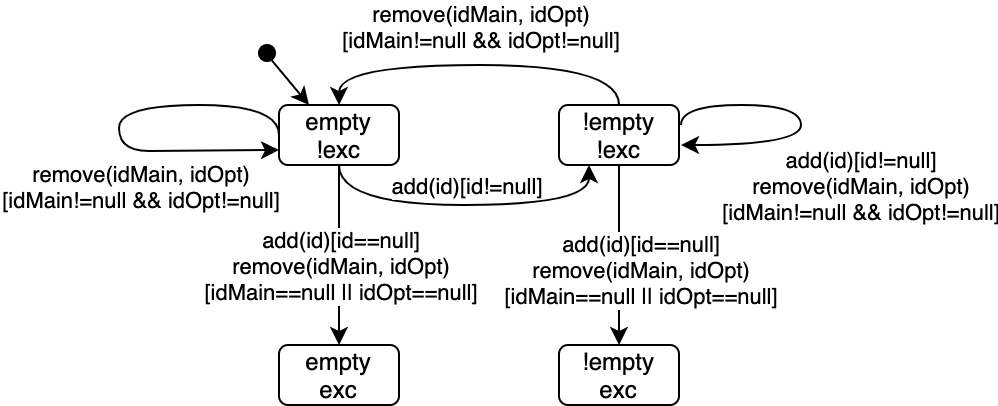}
 \caption{Behavioural model for the \code{TreeSet} implementation}
 \label{fig:v2}
\end{figure}
These models show the result of calling the API methods add and remove. We model the states with the predicates we used to create this abstraction. 
The transitions are labeled with the API method names, some of them have the guards (values in square brackets) which are the conditions based on the methods variables evaluated dynamically, that have to be satisfied in order to fire the transition.

\section{Model extraction approach}
\label{secn:approach}

\subsection{Simplifying assumptions}
\label{secn:assumptions}

We based our work on Java Collection Framework JDK 17\mdash the most recent version at the time of writing~\cite{javacollections}\mdash and considered its General Purpose Implementations (GPI).  We studied each GPI to capture its behaviour and to define its operational semantics taking into account differences in the implementation of methods for various interfaces. For the sake of brevity, we discuss only \code{Set} collections.

We focus on the contribution of collections to the object state.  Thus, the defining property of sets, \ie the absence of duplicates, is irrelevant.  Other properties that can be taken into account are
\begin{textlist}
\item the absence of duplicates, 
\item whether the collection is ordered and 
\item whether it permits null values. 
\end{textlist}
Here, we focus on the latter.

Furthermore, since the operation of enlarging the collection is called implicitly, we disregard the initial capacity and the load factor of the collection.  Similarly, since element retrieval is not based on their position in the collection, we do not have to keep track of the insertion order.  In particular, these assumptions allow us to process \code{HashSet} and \code{LinkedHashSet} in the same manner. For \code{TreeSet}, we have extra rules for adding and removing \code{null} elements.

In addition, we make the following assumptions
that simplify the presentation.  We summarise them here\mdash further details are provided in \secn{encoding}.  We will discuss the validity of these assumptions in \secn{discussion}.  We assume that:

\begin{enumerate}
\item \label{hyp:conditions} all conditions used in branching statements (\code{if}, \code{while} \etc[]) are Boolean expressions referring exclusively to comparisons of values and the \code{contains()} and \code{isEmpty()} operations on collections
\item \label{hyp:formatting} the source code is formatted in such a manner that there is at most one operator per line, all branching statements have all their openning and closing braces and the closing braces are placed on separate lines from any other operators (\cf the example in \fig{example})
\item \label{hyp:types} there are only two types\mdash collections and values\mdash and any value can be stored in any collection
\item \label{hyp:loops} all loops terminate.
\end{enumerate}
We plan to gradually resolve these simplifications up to the limits modern model checking and other static analysis techniques will permit. 


\subsection{Boolean encoding of method behaviour}
\label{secn:encoding}

Our approach relies on the predicate abstraction idea from \cite{DBLP:conf/cav/GrafS97}.  For each method in the API under consideration, we proceed as follows.

Given a method, 
we consider a \emph{context} comprising a set of collection symbols $\Collections$ and a set of value symbols $\Values$.   The symbols in $\Values$ can be literal constants (\eg \code{``Hello!"}, \code{null}), variables, method parameters \etc[] For two symbols $v_1, v_2 \in \Values$ (idem for $c_1, c_2 \in \Collections$), we write $v_1 \equiv v_2$ if they are the same symbol; $v_1 = v_2$ denotes the predicate stating that their values are equal. 

The definition of the context is a parameter of the approach, determining, in particular, the degree of refinement of the resulting model.
Context values are used to form the following predicates:
\begin{itemize}
\item $\equal{v_1}{v_2} \bydef (v_1 = v_2)$, for $v_1, v_2 \in \Values$ ($v_1 \not\equiv v_2$) 
\item $\equal{c_1}{c_2} \bydef (c_1 = c_2)$, for  $c_1, c_2 \in \Collections$ ($c_1 \not\equiv c_2$) 
\item $\contains{c}{v}$ holds for $v \in \Values$ and $c \in \Collections$ if $v$ is contained in $c$
\item $\empt{c}$ holds for $c \in \Collections$ if $c$ is empty
\item $\exception$ holds if an exception has been thrown
\end{itemize}

We postulate the following axioms,
\begin{enumerate}
\item \label{ax:same:value} $\axiom{same:value}(c,v_1,v_2) \bydef (\equal{v_1}{v_2} \implies \contains{c}{v_1} = \contains{c}{v_2})$
\item \label{ax:same:collection} $\axiom{same:collection}(c_1,c_2,v) \bydef (\equal{c_1}{c_2} \implies \contains{c_1}{v} = \contains{c_2}{v})$
\item \label{ax:empty} $\axiom{empty}(c,v) \bydef (\empt{c} \implies \lnot \contains{c}{v})$
\end{enumerate}

Let $\Predicates$ be the set of all the predicates above and $\Predicates' \bydef \setdef{p'}{p \in \Predicates}$.  Assume that all operators of the method are numbered sequentially from 1 to $n$, with the numbering of branching operators corresponding to their closing braces (\cf \fig{example}).  For each predicate $p \in P$ and each $i \in [0,n]$, we define a Boolean variable $p^i$ corresponding to the valuation of $p$ after the execution of the operator $i$.

For a formula $\varphi \in \sB[\Predicates,\Predicates']$, denote
$\varphi^{i,j} \bydef \varphi\bigl[\Predicates^i/\Predicates\bigr]\bigl[\Predicates^j/\Predicates'\bigr]$
the formula where each predicate $p$ (\resp $p'$) is substituted by the variable $p^i$ (\resp $p^j$).  Similarly, $\varphi^i \bydef \varphi[\Predicates^i/\Predicates]$, for $\varphi \in \sB[\Predicates]$.

For a sequence of operators \code{s} and two indices $i, j \in [0,n]$ ($i < j$), we define the encoding function \encoding{s}{i,j} as shown in \fig{encoding}, where \last{s} denotes \emph{the index of the last operator} in a sequence \code{s} and, in \eq{op}, $\varphi_\code{op}$ is the formula encoding the semantics of the operator \code{op} (see \secn{operators} below).  A method \method{} with the body \code{s} is then encoded by the formula
\[
\encoding{\method}{} \bydef \axioms^0 \land \encoding{s}{0, \last{s}}
\,,
\]
with the formula $\axioms$ defined by \eq{axioms}.

\begin{figure}
  \begin{flalign}
    \label{eq:seq}
    &\lefteqn{\encoding{op;s}{i,j} \bydef \encoding{op}{i,\last{op}} \land \encoding{s}{\last{op},j}}
    \\\label{eq:while}
    &\lefteqn{\encoding{while\, (cnd)\ s}{i,j} \bydef \lnot \condition^j}
    \\\nonumber
    &\lefteqn{\encoding{if\, (cnd)\ s_1\ else\ s_2}{i,j} \bydef
      \encoding{s_1}{i,\last{s_1}} \land \encoding{s_2}{i,\last{s_2}}}
    \\\label{eq:if}
    &&{} \land \bigl(\condition^i \implies \bigwedge_{p \in \Predicates} p^j = p^{\last{s_1}}\bigr)
    \land \bigl(\lnot \condition^i \implies \bigwedge_{p \in \Predicates} p^j = p^{\last{s_2}}\bigr)
    \\\label{eq:op}
    &\lefteqn{\encoding{op}{i,j} \bydef \varphi_{\code{op}}^{i,j} \land \axioms^j}
    \\\label{eq:axioms}
    &\lefteqn{\axioms \bydef {}}
    \\\nonumber
    &&\bigwedge_{\rangestackiii{c \in \Collections}{v_1, v_2 \in \Values}{v_1 \not\equiv v_2}}
      \axiom{same:value}(c,v_1,v_2)
      \land
      \bigwedge_{\rangestackiii{c_1, c_2 \in \Collections}{v \in \Values}{c_1 \not\equiv c_2}}
      \axiom{same:collection}(c_1,c_2,v)
      \land
      \bigwedge_{\rangestackii{c \in \Collections}{v \in \Values}}
      \axiom{empty}(c,v)
  \end{flalign}
  \caption{Definition of the encoding function \encoding{\cdot}{\cdot,\cdot}
  }
  \label{fig:encoding}
\end{figure}


\subsection{Predicate semantics of operators}
\label{secn:operators}

The predicate semantics of an operator \code{op} is given by a quadruple $(\Collections[\code{op}], \Values[\code{op}], \opset{\code{op}}, \opformula{\code{op}})$, where
\begin{itemize}
\item \Context[\code{op}] is a context, comprising the sets $\Collections[\code{op}]$ and $\Values[\code{op}]$ of, respectively, collection and value variables
\item $\opset{\code{op}}$ is a subset of predicates as in \secn{encoding} indexed by the variables from $\Collections[\code{op}]$ and $\Values[\code{op}]$
\item $\opformula{\code{op}} \in \sB[\opset{\code{op}}, \opset{\code{op}}']$ is a Boolean formula on these predicates and $\sB[\opset{\code{op}}, \opset{\code{op}}']$ is Boolean algebra generated by $\opset{\code{op}}\bigcup \opset{\code{op}}'$ (set of Boolean formulas with variables from P and P’)
\end{itemize}
Intuitively, $\opset{\code{op}}$ is the set of predicates whereof the valuations \emph{may} be affected by the operator.  

For example, consider the method \code{c.add(v)} of a \code{HashSet} \code{c}.  Its predicate semantics is given by the formula
\[
\opformula{\code{c.add(v)}} \bydef \contains{c}{v}' \land \lnot\empt{c}'
\,.
\]
Here, $\Collections[\code{c.add(v)}] = \{c\}$, $\Values[\code{c.add(v)}] = \{v\}$, $\opset{\code{c.add(v)}} = \{\contains{c}{v}, \empt{c}\}$.

Let us consider, as another example, the method \code{c.clear()} of a collection \code{c}.  Its predicate semantics is given by the formula
\[
\opformula{\code{c.clear()}} \bydef \empt{c}'
\,.
\]
Despite $v$ not appearing in \opformula{\code{c.clear()}}, we take the same $\Collections[\code{c.clear()}] = \{c\}$,  $\Values[\code{c.clear()}] = \{v\}$ and $\opset{\code{c.clear()}} = \{\contains{c}{v}, \empt{c}\}$ as above.  Indeed, the value of the \contains{c}{v} predicate, \emph{for all value symbols in the context}, may be modified by this method, even though $\axiom{empty}$ means that this does not have to be specified explicitly in \opformula{\code{c.clear()}}.

For each operator invocation in the source code, some of the variables in $\Collections[\code{op}]$ and $\Values[\code{op}]$ will be positionally matched with the symbols in the context $\Context$.  \Eg[,] for an invocation \code{idSet.clear()}, $c \in \Collections[\code{c.clear()}]$ is matched to $\mathit{idSet} \in \Collections$ but $v \in \Values[\code{c.clear()}]$ is not matched to any symbol in $\Values$.  For the sake of clarity, let us abuse the notation as follows: we denote 
$\Collections[\code{op}] \setminus \Collections$ (\resp $\Values[\code{op}] \setminus \Values$) the set of variables that are not matched positionally.

The formula $\varphi_{\code{op}}$ encoding the operator \code{op} (see \eq{op} in \fig{encoding}) is then defined as follows:
\begin{gather}
  \label{eq:operator:encoding}
  \varphi_{\code{op}} \bydef
  \bigwedge_{\rangestackii{\mu : \Collections[\code{op}] \setminus \Collections \rightarrow \Collections}{\nu : \Values[\code{op}] \setminus \Values \rightarrow \Values}}
  \opformula{\code{op}}[\mu,\nu]
  \land
  \bigwedge_{p \in \free{\Predicates}} p' = p
  \\\nonumber
  \text{with} \qquad
  \free{\Predicates} \bydef
  \Predicates \setminus
  \bigcup_{\rangestackii{\mu : \Collections[\code{op}] \setminus \Collections \rightarrow \Collections}{\nu : \Values[\code{op}] \setminus \Values \rightarrow \Values}}
  \opset{\code{op}}[\mu,\nu]
  \,,
\end{gather}
where $\opformula{\code{op}}[\mu,\nu]$ denotes the formula obtained by substituting each unmatched variable $c \in \Collections[\code{op}] \setminus \Collections$ (\resp $v \in \Values[\code{op}] \setminus \Values$) by the symbol $\mu(c)$ (\resp $\nu(v)$) and similarly for $\opset{\code{op}}[\mu,\nu]$.

Intuitively, the first conjunct in \eq{operator:encoding} means that the formula \opformula{\code{op}} is instantiated for all possible values of the unmatched variables.  The second conjunct means that all predicates in $\Predicates$ that do not appear in the semantics of \code{op} maintain their previous values.

Notice, finally, that a predicate $p'$, with $p \in \opset{\code{op}}$, that is not fully constrained by \opformula{\code{op}} might take any value, representing the absence of the corresponding information.  For instance, consider the method \code{c.remove(v)} of a \code{HashSet} \code{c}.  Its predicate semantics is given by the formula
\[
\opformula{\code{c.remove(v)}} \bydef
\lnot \contains{c}{v}' \land (\empt{c} \implies \empt{c}')
\,,
\]
with, again, $\Collections[\code{c.remove(v)}] = \{c\}$,  $\Values[\code{c.remove(v)}] = \{v\}$ and $\opset{\code{c.remove(v)}} = \{\contains{c}{v}, \empt{c}\}$.  If \code{c} is not empty before the call to \code{c.remove(v)} it may or may not become empty after the call, depending on whether \code{v} is the only value contained therein.


\subsection{Computing the FSM}
\label{secn:fsm}

Given a context $\Context$ with the corresponding set of predicates $\Predicates$ as defined in \secn{encoding}, we observe that the symbols can be partitioned into \emph{state} and \emph{indeterminacy} symbols: $\Collections = \Collections[st] \uplus \Collections[nd]$ and $\Values = \Values[st] \uplus \Values[nd]$, such that \Collections[st] and \Values[st] contain constants and class fields, whereas \Collections[nd] and \Values[nd] contain method parameters, unevaluated functions \etc  This induces several partitions on the set of predicates: $\Predicates = \Predicates[st] \uplus \Predicates[nd]$, such that \Predicates[st] is a subset of predicates that refer only to symbols in \Collections[st] and \Values[st]. 

A \emph{state} is a partial valuation $\State: \Predicates[st] \rightharpoonup \sB$.  A \emph{concrete state} is a total valuation $\State: \Predicates[st] \rightarrow \sB$.  We say that one state $\State[1]$  \emph{refines} another $\State[2]$, denoted \refines{\State[1]}{\State[2]} if $\domain{\State[2]} \subseteq \domain{\State[1]}$ and $\State[1](p) = \State[2](p)$, for all $p \in \domain{\State[2]}$.  We say that a set of states $\States$ is \emph{complete} if every state refines some state of $\States$, \ie for every $\State: \Predicates[st] \rightharpoonup \sB$, there exists $\hat{\State} \in \States$, such that \refines{\State}{\hat{\State}}.
Notice that any complete set of states must comprise the empty valuation $\State[0]$, having $\domain{\State[0]}=\emptyset$.  Clearly, for any given implementation of an API, there can be different complete sets of states.  In particular, the set $\{\State[0]\}$ is complete.  So is the set of all concrete states $\bsetdef{\State}{\domain{\State}=\Predicates[st]} \cup \{\State[0]\}$.

Given a context $\Context$, a set of predicates $\Predicates$ and a complete set of states $\States$, we build a non-deterministic FSM $(\States, \State[0], \api, \goesto{})$, where \State[0] is the initial state, $\api$ is the set of methods provided by the API under consideration and ${\goesto{}} \subseteq \States \times \api \times \sB[\Predicates[nd]] \times \States$ is the transition relation,
where the third component is the transition \emph{guard}.

To compute the transition relation \goesto{}, we proceed as follows.  For each method $\method \in \api$ and each pair of states $\State[1], \State[2] \in \States$, we define
\[
\varphi_0 \bydef
\bigwedge_{p \in \dom{\State[1]}} \bigl(p^0 = \State[1](p)\bigr)
\land
\encoding{\method}{}
\land
\bigwedge_{p \in \dom{\State[2]}} \bigl(p^{\last{\method}} = \State[2](p)\bigr)
\]
(see \secn{encoding}) and submit it to a SAT solver.  If $\varphi_0$ is satisfiable we obtain a model $m_1: \setdef{p^i}{p \in \Predicates, i \in [0,\last{\method}]} \rightarrow \sB$.  We define
\[
\varphi_1 \bydef
\varphi_0 \land
\lnot\bigwedge_{i=0}^{\last{\method}}\bigwedge_{p \in \Predicates}
\bigl(p^i = m_1(p^i)\bigr)
\]
and submit it to the SAT solver again, repeating until we obtain some $\varphi_n$ that is unsatisfiable.  The transition is then
\[
\State[1] \goesto{\method[g]} \State[2]\,,
\qquad\text{with }
g = \bigvee_{i = 1}^n \bigwedge_{p \in \Predicates[nd]}
\bigl(p = m_i(p^0)\bigr)
\,.
\]
Notice that, if $\varphi_0$ is unsatisfiable, we have $g = \false$.

Applying this approach to the example in \fig{example} we obtain the FSMs in \figs{v1}{v2}.  
We do so using the set of all concrete states, in which each state is a combination of predicates from $\Predicates$ (predicates forming the states could be selected to tune the level of the model's abstraction, e.g in our example we rely on the $\empt{c}$ predicate even though $\Predicates$ is not limited by it.
The common context for the \code{HashSet} version of the example is $(\{\mathit{idSet}\},\emptyset)$.  For methods \code{add} and \code{removeId} we take the union with, respectively, $(\emptyset, \{\mathit{id}\})$ and $(\emptyset, \{\mathit{idMain}, \mathit{idOpt}\})$.  The predicates are the same as in \secn{encoding}, with $\Predicates[st] = \{\empt{\mathit{idSet}}\}$ (we exclude \exception, since it appears neither in the axioms nor in the semantics of the operators involved).

For the \code{TreeSet} version, we add $\mathit{null}$ to the common context and \exception{} to \Predicates[st] since they appear in the semantics of the operators, \eg
\begin{multline*}
  \opformula{\code{c.add(v)}} \bydef 
  (\equal{v}{\mathit{null}} = \equal{v}{\mathit{null}}')
  \\
  {} \land
  \lnot\equal{v}{\mathit{null}} \implies (\contains{c}{v}' \land \lnot \empt{c}' \land (\exception' = \exception))
  \\
  {} \land
  \equal{v}{\mathit{null}} \implies (\contains{c}{v} = \contains{c}{v}' \land \empt{c} = \empt{c}' \land \exception')
\end{multline*}
\begin{multline*}
  \opformula{\code{c.remove(v)}} \bydef
  (\equal{v}{\mathit{null}} = \equal{v}{\mathit{null}}')
  \\
  {} \land
  \lnot\equal{v}{\mathit{null}} \implies (\lnot \contains{c}{v}' \land (\empt{c} \implies \empt{c}') \land (\exception' = \exception))
  \\
  {} \land
  \equal{v}{\mathit{null}} \implies (\contains{c}{v} = \contains{c}{v}' \land \empt{c} = \empt{c}' \land \exception')
\end{multline*}

The full Z3 encodings of the two versions can be found in \cite{anoymous_2022_5913271}.


\section{Discussion \& Future work}
\label{secn:discussion}

The model extraction approach proposed above is based on ideas that are well-known in the Formal Methods community.  The novelty lies in their application for the extraction of behavioural models from Java source code.  More importantly, our main goal is not to use these models for verification but rather for assisting software engineers in the development process, making API more usable due to the provision of additional information on the expected order of the calls of the operation, and, eventually, for model-based monitoring and coordination. This allows us, in particular, to aim for a less restrictive behavioural model, namely non-deterministic FSMs with guards.  This change of perspective also makes the semi-automatic approach more acceptable: indeed, we believe that it is much easier to get developers to help refining an existing model than to provide one from scratch. Both limitations and the advantages of the approach refer to the problem of building the model with the right level of abstraction. Abstracting excessively leads to primitive models, useless in a practical context. Abstracting insufficiently leads to undecidability and the state explosion problem. We leave implementation of abstraction-tuning possibility as a future work.

The semi-automatic nature of our approach arises from several levels of parametrisation among which the key one is choosing the appropriate predicates for the specification of the semantics of the basic operations.  To some extent, this remark addresses the simplifying assumption \hyp{conditions} in \secn{assumptions}, which can be substantially relaxed by including additional predicates to define the semantics of the terms used in the conditions.  Further relaxation can, of course, be achieved by moving from SAT to SMT.

Simplifying hypothesis do not  prevent aliasing, but alias discovery is currently out of our scope. However, if we know that one variable is aliasing the other one, we can force the corresponding equality predicate for them (set it to true). Otherwise the alias problem would be resolved dynamically since the equality predicate will appear in transition guards.

Deciding which symbols should be part of the context and which predicates should be used for the encoding is a non-trivial task, which we believe cannot be fully formalised.  We see two approaches to address this question: input from domain experts, \ie developers, and various kinds of heuristics (potentially including Machine Learning techniques).  Both approaches\mdash or combinations whereof\mdash provides directions for future work.

Assumptions \hyp{formatting} and \hyp{types} are obviously non-constraining.  Moreover, in any actual implementation, we would use Abstract Syntax Trees and Control Flow Graphs to explore the source code and compute the Boolean encodings.  This would naturally eliminate the need for these two assumptions.

Thus, the strongest assumption that we have made is the loop termination assumption \hyp{loops}, which would require deeper static analysis or assistance from the developers for exhibiting loop variants, which we consider beyond the near-future scope of this work.

\begin{figure}
 \centering
 \includegraphics[width=0.45\textwidth]{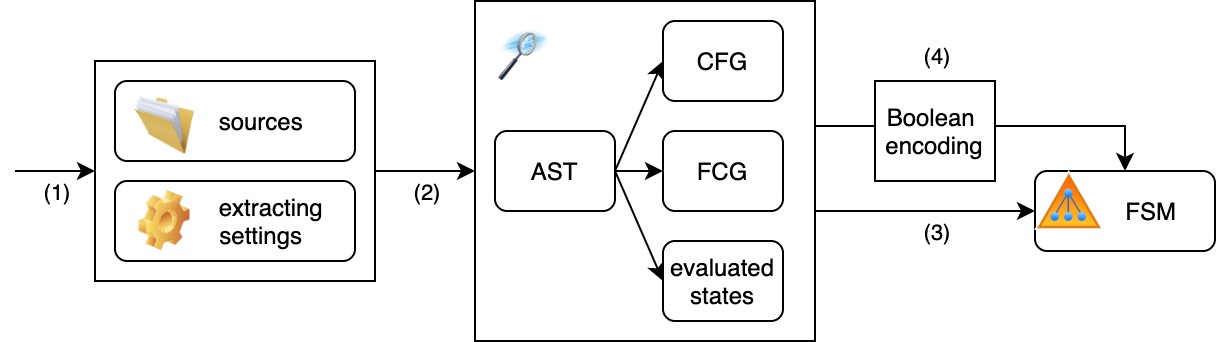}
 \caption{Intended tool architecture}
 \label{fig:arch}
\end{figure}

We intend to develop a tool for extracting FSMs from Java source code based on the approach proposed in this paper. 
In general, the tool will proceed along the following steps (see \fig{arch}): collecting the information about the sources under analysis and about the fields identified as key to the object state~(1); computing auxiliary data structures, and state values that can be easily evaluated~(2); then using different techniques to generate the FSM. The last step differs for the basic types and collections.  For instance, for simple types, such as enumerations, simple interpretative techniques could suffice~(3).  For complex types such as collections, we will rely on the SAT-based approach presented in this paper~(4). 

Finally, another point worth mentioning is the choice of the complete set of states.  For the sake of simplicity, we have chosen to use the set of all concrete states.  This can clearly be improved both a posteriori\mdash \eg reducing the FSM by bisimilarity\mdash and a priori\mdash \eg building the state space on the fly to avoid unreachable states.


\section{Conclusion}
\label{secn:conclusion}
We propose an approach to the extraction of behavioural models in the form of non-deterministic FSMs with guards from Java implementations of library APIs.  The approach is semi-automatic since it is parametrised by the choice of important symbols, predicates and states.  This makes it highly flexible and allows the integration of expert knowledge from developers.  The presented approach is fully formalised.  Although the underlying ideas are well known in the Formal Methods community, we believe that their combined application for the extraction of FSMs with guards is novel and can be applied to the benefit of the Software Engineering community.  We are planning to develop a tool based on the proposed approach and use it in our future projects for the purposes of monitoring and coordination of software components.

\bibliographystyle{ACM-Reference-Format}
\bibliography{bibliography}


\begin{thebibliography}{30}


\ifx \showCODEN    \undefined \def \showCODEN     #1{\unskip}     \fi
\ifx \showDOI      \undefined \def \showDOI       #1{#1}\fi
\ifx \showISBNx    \undefined \def \showISBNx     #1{\unskip}     \fi
\ifx \showISBNxiii \undefined \def \showISBNxiii  #1{\unskip}     \fi
\ifx \showISSN     \undefined \def \showISSN      #1{\unskip}     \fi
\ifx \showLCCN     \undefined \def \showLCCN      #1{\unskip}     \fi
\ifx \shownote     \undefined \def \shownote      #1{#1}          \fi
\ifx \showarticletitle \undefined \def \showarticletitle #1{#1}   \fi
\ifx \showURL      \undefined \def \showURL       {\relax}        \fi
\providecommand\bibfield[2]{#2}
\providecommand\bibinfo[2]{#2}
\providecommand\natexlab[1]{#1}
\providecommand\showeprint[2][]{arXiv:#2}

\bibitem[\protect\citeauthoryear{??}{jav}{2021}]%
        {javacollections}
 \bibinfo{year}{2021}\natexlab{}.
\newblock \bibinfo{title}{{Java Collections Framework Overview}}.
\newblock
  \bibinfo{howpublished}{\url{https://docs.oracle.com/en/java/javase/17/docs/api/java.base/java/util/doc-files/coll-overview.html}}.
\newblock
\newblock
\shownote{[Online; accessed 6-Jan-2022]}.


\bibitem[\protect\citeauthoryear{Ancona, Bono, Bravetti, Campos, Castagna,
  Deni{\'{e}}lou, Gay, Gesbert, Giachino, Hu, Johnsen, Martins, Mascardi,
  Montesi, Neykova, Ng, Padovani, Vasconcelos, and Yoshida}{Ancona
  et~al\mbox{.}}{2016}]%
        {DBLP:journals/ftpl/AnconaBB0CDGGGH16}
\bibfield{author}{\bibinfo{person}{Davide Ancona}, \bibinfo{person}{Viviana
  Bono}, \bibinfo{person}{Mario Bravetti}, \bibinfo{person}{Joana Campos},
  \bibinfo{person}{Giuseppe Castagna}, \bibinfo{person}{Pierre{-}Malo
  Deni{\'{e}}lou}, \bibinfo{person}{Simon~J. Gay}, \bibinfo{person}{Nils
  Gesbert}, \bibinfo{person}{Elena Giachino}, \bibinfo{person}{Raymond Hu},
  \bibinfo{person}{Einar~Broch Johnsen}, \bibinfo{person}{Francisco Martins},
  \bibinfo{person}{Viviana Mascardi}, \bibinfo{person}{Fabrizio Montesi},
  \bibinfo{person}{Rumyana Neykova}, \bibinfo{person}{Nicholas Ng},
  \bibinfo{person}{Luca Padovani}, \bibinfo{person}{Vasco~T. Vasconcelos},
  {and} \bibinfo{person}{Nobuko Yoshida}.} \bibinfo{year}{2016}\natexlab{}.
\newblock \showarticletitle{Behavioral Types in Programming Languages}.
\newblock \bibinfo{journal}{\emph{Found. Trends Program. Lang.}}
  \bibinfo{volume}{3}, \bibinfo{number}{2-3} (\bibinfo{year}{2016}),
  \bibinfo{pages}{95--230}.
\newblock
\urldef\tempurl%
\url{https://doi.org/10.1561/2500000031}
\showDOI{\tempurl}


\bibitem[\protect\citeauthoryear{Ancona, Dagnino, and Franceschini}{Ancona
  et~al\mbox{.}}{2018}]%
        {ancona18}
\bibfield{author}{\bibinfo{person}{Davide Ancona}, \bibinfo{person}{Francesco
  Dagnino}, {and} \bibinfo{person}{Luca Franceschini}.}
  \bibinfo{year}{2018}\natexlab{}.
\newblock \showarticletitle{A Formalism for Specification of Java API
  Interfaces}. In \bibinfo{booktitle}{\emph{Companion Proceedings for the
  ISSTA/ECOOP 2018 Workshops}} (Amsterdam, Netherlands)
  \emph{(\bibinfo{series}{ISSTA '18})}. \bibinfo{publisher}{Association for
  Computing Machinery}, \bibinfo{address}{New York, NY, USA},
  \bibinfo{pages}{24–26}.
\newblock
\showISBNx{9781450359399}
\urldef\tempurl%
\url{https://doi.org/10.1145/3236454.3236476}
\showDOI{\tempurl}


\bibitem[\protect\citeauthoryear{Anoymous}{Anoymous}{2022}]%
        {anoymous_2022_5913271}
\bibfield{author}{\bibinfo{person}{Anoymous}.} \bibinfo{year}{2022}\natexlab{}.
\newblock \bibinfo{title}{Paper's example z3-encoding}.
\newblock
\newblock
\urldef\tempurl%
\url{https://doi.org/10.5281/zenodo.5913271}
\showDOI{\tempurl}


\bibitem[\protect\citeauthoryear{Bartolo~Burlò, Francalanza, and
  Scalas}{Bartolo~Burlò et~al\mbox{.}}{2020}]%
        {burlo20}
\bibfield{author}{\bibinfo{person}{Christian Bartolo~Burlò},
  \bibinfo{person}{Adrian Francalanza}, {and} \bibinfo{person}{Alceste
  Scalas}.} \bibinfo{year}{2020}\natexlab{}.
\newblock \bibinfo{booktitle}{\emph{Towards a Hybrid Verification Methodology
  for Communication Protocols (Short Paper)}}.
\newblock \bibinfo{pages}{227--235}.
\newblock
\showISBNx{978-3-030-50085-6}
\urldef\tempurl%
\url{https://doi.org/10.1007/978-3-030-50086-3_13}
\showDOI{\tempurl}


\bibitem[\protect\citeauthoryear{Bloch}{Bloch}{2006}]%
        {bloch06}
\bibfield{author}{\bibinfo{person}{Joshua Bloch}.}
  \bibinfo{year}{2006}\natexlab{}.
\newblock \showarticletitle{How to Design a Good API and Why It Matters}. In
  \bibinfo{booktitle}{\emph{Companion to the 21st ACM SIGPLAN Symposium on
  Object-Oriented Programming Systems, Languages, and Applications}} (Portland,
  Oregon, USA) \emph{(\bibinfo{series}{OOPSLA '06})}.
  \bibinfo{publisher}{Association for Computing Machinery},
  \bibinfo{address}{New York, NY, USA}, \bibinfo{pages}{506–507}.
\newblock
\showISBNx{159593491X}
\urldef\tempurl%
\url{https://doi.org/10.1145/1176617.1176622}
\showDOI{\tempurl}


\bibitem[\protect\citeauthoryear{Bravetti, Giallorenzo, Mauro, Talevi, and
  Zavattaro}{Bravetti et~al\mbox{.}}{2020}]%
        {Bravetti2020AFA}
\bibfield{author}{\bibinfo{person}{Mario Bravetti}, \bibinfo{person}{Saverio
  Giallorenzo}, \bibinfo{person}{Jacopo Mauro}, \bibinfo{person}{Iacopo
  Talevi}, {and} \bibinfo{person}{Gianluigi Zavattaro}.}
  \bibinfo{year}{2020}\natexlab{}.
\newblock \showarticletitle{A Formal Approach to Microservice Architecture
  Deployment}. In \bibinfo{booktitle}{\emph{Microservices, Science and
  Engineering}}.
\newblock


\bibitem[\protect\citeauthoryear{Bravetti and Zavattaro}{Bravetti and
  Zavattaro}{2019}]%
        {Bravetti19}
\bibfield{author}{\bibinfo{person}{Mario Bravetti} {and}
  \bibinfo{person}{Gianluigi Zavattaro}.} \bibinfo{year}{2019}\natexlab{}.
\newblock \bibinfo{booktitle}{\emph{Relating Session Types and Behavioural
  Contracts: The Asynchronous Case}}.
\newblock \bibinfo{pages}{29--47}.
\newblock
\showISBNx{978-3-030-30445-4}
\urldef\tempurl%
\url{https://doi.org/10.1007/978-3-030-30446-1_2}
\showDOI{\tempurl}


\bibitem[\protect\citeauthoryear{Brito, Xavier, Hora, and Valente}{Brito
  et~al\mbox{.}}{2018}]%
        {brito18}
\bibfield{author}{\bibinfo{person}{Aline Brito}, \bibinfo{person}{Laerte
  Xavier}, \bibinfo{person}{Andre Hora}, {and} \bibinfo{person}{Marco
  Valente}.} \bibinfo{year}{2018}\natexlab{}.
\newblock \showarticletitle{Why and how Java developers break APIs}.
  \bibinfo{pages}{255--265}.
\newblock
\urldef\tempurl%
\url{https://doi.org/10.1109/SANER.2018.8330214}
\showDOI{\tempurl}


\bibitem[\protect\citeauthoryear{Brito, Hora, Valente, and Robbes}{Brito
  et~al\mbox{.}}{2017}]%
        {brito17}
\bibfield{author}{\bibinfo{person}{Gleison Brito}, \bibinfo{person}{Andre
  Hora}, \bibinfo{person}{Marco Valente}, {and} \bibinfo{person}{Romain
  Robbes}.} \bibinfo{year}{2017}\natexlab{}.
\newblock \showarticletitle{On the Use of Replacement Messages in API
  Deprecation: An Empirical Study}.
\newblock \bibinfo{journal}{\emph{Journal of Systems and Software}}
  \bibinfo{volume}{137} (\bibinfo{date}{12} \bibinfo{year}{2017}).
\newblock
\urldef\tempurl%
\url{https://doi.org/10.1016/j.jss.2017.12.007}
\showDOI{\tempurl}


\bibitem[\protect\citeauthoryear{Cook, Bock, Rivett, Rutt, Seidewitz, Selic,
  and Tolbert}{Cook et~al\mbox{.}}{2017}]%
        {Cook2017}
\bibfield{author}{\bibinfo{person}{Steve Cook}, \bibinfo{person}{Conrad Bock},
  \bibinfo{person}{Pete Rivett}, \bibinfo{person}{Tom Rutt},
  \bibinfo{person}{Ed Seidewitz}, \bibinfo{person}{Bran Selic}, {and}
  \bibinfo{person}{Doug Tolbert}.} \bibinfo{year}{2017}\natexlab{}.
\newblock \bibinfo{booktitle}{\emph{Unified Modeling Language ({UML}) Version
  2.5.1}}.
\newblock \bibinfo{type}{Standard}. \bibinfo{institution}{Object Management
  Group ({OMG})}.
\newblock
\urldef\tempurl%
\url{https://www.omg.org/spec/UML/2.5.1}
\showURL{%
\tempurl}


\bibitem[\protect\citeauthoryear{Corbett}{Corbett}{2000}]%
        {10.1145/332740.332741}
\bibfield{author}{\bibinfo{person}{James~C. Corbett}.}
  \bibinfo{year}{2000}\natexlab{}.
\newblock \showarticletitle{Using Shape Analysis to Reduce Finite-State Models
  of Concurrent Java Programs}.
\newblock \bibinfo{journal}{\emph{ACM Trans. Softw. Eng. Methodol.}}
  \bibinfo{volume}{9}, \bibinfo{number}{1} (\bibinfo{date}{Jan.}
  \bibinfo{year}{2000}), \bibinfo{pages}{51–93}.
\newblock
\showISSN{1049-331X}
\urldef\tempurl%
\url{https://doi.org/10.1145/332740.332741}
\showDOI{\tempurl}


\bibitem[\protect\citeauthoryear{Cruz-Filipe, Larsen, and Montesi}{Cruz-Filipe
  et~al\mbox{.}}{2017}]%
        {luis17}
\bibfield{author}{\bibinfo{person}{Lu{\'i}s Cruz-Filipe},
  \bibinfo{person}{Kim~S. Larsen}, {and} \bibinfo{person}{Fabrizio Montesi}.}
  \bibinfo{year}{2017}\natexlab{}.
\newblock \showarticletitle{The Paths to Choreography Extraction}. In
  \bibinfo{booktitle}{\emph{Foundations of Software Science and Computation
  Structures}}, \bibfield{editor}{\bibinfo{person}{Javier Esparza} {and}
  \bibinfo{person}{Andrzej~S. Murawski}} (Eds.). \bibinfo{publisher}{Springer
  Berlin Heidelberg}, \bibinfo{address}{Berlin, Heidelberg},
  \bibinfo{pages}{424--440}.
\newblock


\bibitem[\protect\citeauthoryear{Dardha, Gay, Kouzapas, Perera, Voinea, and
  Weber}{Dardha et~al\mbox{.}}{2017}]%
        {dardha2017mungo}
\bibfield{author}{\bibinfo{person}{Ornela Dardha}, \bibinfo{person}{Simon~J
  Gay}, \bibinfo{person}{Dimitrios Kouzapas}, \bibinfo{person}{Roly Perera},
  \bibinfo{person}{A~Laura Voinea}, {and} \bibinfo{person}{Florian Weber}.}
  \bibinfo{year}{2017}\natexlab{}.
\newblock \showarticletitle{Mungo and StMungo: Tools for Typechecking Protocols
  in Java}.
\newblock \bibinfo{journal}{\emph{Behavioural Types: from Theory to Tools}}
  (\bibinfo{year}{2017}), \bibinfo{pages}{309}.
\newblock


\bibitem[\protect\citeauthoryear{Dietrich, Jezek, and Brada}{Dietrich
  et~al\mbox{.}}{2014}]%
        {dietrich14}
\bibfield{author}{\bibinfo{person}{Jens Dietrich}, \bibinfo{person}{Kamil
  Jezek}, {and} \bibinfo{person}{Premek Brada}.}
  \bibinfo{year}{2014}\natexlab{}.
\newblock \showarticletitle{Broken promises: An empirical study into evolution
  problems in Java programs caused by library upgrades}. In
  \bibinfo{booktitle}{\emph{2014 Software Evolution Week - IEEE Conference on
  Software Maintenance, Reengineering, and Reverse Engineering (CSMR-WCRE)}}.
  \bibinfo{pages}{64--73}.
\newblock
\urldef\tempurl%
\url{https://doi.org/10.1109/CSMR-WCRE.2014.6747226}
\showDOI{\tempurl}


\bibitem[\protect\citeauthoryear{Dig, Comertoglu, Marinov, and Johnson}{Dig
  et~al\mbox{.}}{2006}]%
        {dig06}
\bibfield{author}{\bibinfo{person}{Danny Dig}, \bibinfo{person}{Can
  Comertoglu}, \bibinfo{person}{Darko Marinov}, {and} \bibinfo{person}{Ralph
  Johnson}.} \bibinfo{year}{2006}\natexlab{}.
\newblock \showarticletitle{Automated Detection of Refactorings in Evolving
  Components}.
\newblock
\showISBNx{978-3-540-35726-1}
\urldef\tempurl%
\url{https://doi.org/10.1007/11785477_24}
\showDOI{\tempurl}


\bibitem[\protect\citeauthoryear{Espinha, Zaidman, and Gro{\ss}}{Espinha
  et~al\mbox{.}}{2015}]%
        {Espinha2015WebAG}
\bibfield{author}{\bibinfo{person}{Tiago Espinha}, \bibinfo{person}{Andy
  Zaidman}, {and} \bibinfo{person}{Hans-Gerhard Gro{\ss}}.}
  \bibinfo{year}{2015}\natexlab{}.
\newblock \showarticletitle{Web API growing pains: Loosely coupled yet strongly
  tied}.
\newblock \bibinfo{journal}{\emph{J. Syst. Softw.}}  \bibinfo{volume}{100}
  (\bibinfo{year}{2015}), \bibinfo{pages}{27--43}.
\newblock


\bibitem[\protect\citeauthoryear{Gabbrielli, Giallorenzo, Lanese, and
  Mauro}{Gabbrielli et~al\mbox{.}}{2019}]%
        {GabbrielliGLM19}
\bibfield{author}{\bibinfo{person}{Maurizio Gabbrielli},
  \bibinfo{person}{Saverio Giallorenzo}, \bibinfo{person}{Ivan Lanese}, {and}
  \bibinfo{person}{Jacopo Mauro}.} \bibinfo{year}{2019}\natexlab{}.
\newblock \showarticletitle{Guess Who's Coming: Runtime Inclusion of
  Participants in Choreographies}. In \bibinfo{booktitle}{\emph{The Art of
  Modelling Computational Systems: {A} Journey from Logic and Concurrency to
  Security and Privacy - Essays Dedicated to Catuscia Palamidessi on the
  Occasion of Her 60th Birthday}} \emph{(\bibinfo{series}{Lecture Notes in
  Computer Science}, Vol.~\bibinfo{volume}{11760})},
  \bibfield{editor}{\bibinfo{person}{M{\'{a}}rio~S. Alvim},
  \bibinfo{person}{Kostas Chatzikokolakis}, \bibinfo{person}{Carlos Olarte},
  {and} \bibinfo{person}{Frank Valencia}} (Eds.).
  \bibinfo{publisher}{Springer}, \bibinfo{pages}{118--138}.
\newblock
\urldef\tempurl%
\url{https://doi.org/10.1007/978-3-030-31175-9\_8}
\showDOI{\tempurl}


\bibitem[\protect\citeauthoryear{Godoy, Galeotti, Garbervetsky, and
  Uchitel}{Godoy et~al\mbox{.}}{2021}]%
        {Godoy21}
\bibfield{author}{\bibinfo{person}{Javier Godoy}, \bibinfo{person}{Juan~Pablo
  Galeotti}, \bibinfo{person}{Diego Garbervetsky}, {and}
  \bibinfo{person}{Sebasti\'{a}n Uchitel}.} \bibinfo{year}{2021}\natexlab{}.
\newblock \showarticletitle{Enabledness-Based Testing of Object Protocols}.
\newblock \bibinfo{journal}{\emph{ACM Trans. Softw. Eng. Methodol.}}
  \bibinfo{volume}{30}, \bibinfo{number}{2}, Article \bibinfo{articleno}{12}
  (\bibinfo{date}{jan} \bibinfo{year}{2021}), \bibinfo{numpages}{36}~pages.
\newblock
\showISSN{1049-331X}
\urldef\tempurl%
\url{https://doi.org/10.1145/3415153}
\showDOI{\tempurl}


\bibitem[\protect\citeauthoryear{Graf and Sa{\"{\i}}di}{Graf and
  Sa{\"{\i}}di}{1997}]%
        {DBLP:conf/cav/GrafS97}
\bibfield{author}{\bibinfo{person}{Susanne Graf} {and} \bibinfo{person}{Hassen
  Sa{\"{\i}}di}.} \bibinfo{year}{1997}\natexlab{}.
\newblock \showarticletitle{Construction of Abstract State Graphs with {PVS}}.
  In \bibinfo{booktitle}{\emph{Computer Aided Verification, 9th International
  Conference, {CAV} '97, Haifa, Israel, June 22-25, 1997, Proceedings}}
  \emph{(\bibinfo{series}{Lecture Notes in Computer Science},
  Vol.~\bibinfo{volume}{1254})}, \bibfield{editor}{\bibinfo{person}{Orna
  Grumberg}} (Ed.). \bibinfo{publisher}{Springer}, \bibinfo{pages}{72--83}.
\newblock
\urldef\tempurl%
\url{https://doi.org/10.1007/3-540-63166-6\_10}
\showDOI{\tempurl}


\bibitem[\protect\citeauthoryear{Harel and Gery}{Harel and Gery}{1997}]%
        {harel97}
\bibfield{author}{\bibinfo{person}{D. Harel} {and} \bibinfo{person}{E. Gery}.}
  \bibinfo{year}{1997}\natexlab{}.
\newblock \showarticletitle{Executable object modeling with statecharts}.
\newblock \bibinfo{journal}{\emph{Computer}} \bibinfo{volume}{30},
  \bibinfo{number}{7} (\bibinfo{year}{1997}), \bibinfo{pages}{31--42}.
\newblock
\urldef\tempurl%
\url{https://doi.org/10.1109/2.596624}
\showDOI{\tempurl}


\bibitem[\protect\citeauthoryear{Henning}{Henning}{2007}]%
        {Henning07}
\bibfield{author}{\bibinfo{person}{Michi Henning}.}
  \bibinfo{year}{2007}\natexlab{}.
\newblock \showarticletitle{API: Design Matters: Why Changing APIs Might Become
  a Criminal Offense.}
\newblock \bibinfo{journal}{\emph{Queue}} \bibinfo{volume}{5},
  \bibinfo{number}{4} (\bibinfo{date}{may} \bibinfo{year}{2007}),
  \bibinfo{pages}{24–36}.
\newblock
\showISSN{1542-7730}
\urldef\tempurl%
\url{https://doi.org/10.1145/1255421.1255422}
\showDOI{\tempurl}


\bibitem[\protect\citeauthoryear{Henzinger, Jhala, and Majumdar}{Henzinger
  et~al\mbox{.}}{2005}]%
        {henzinger05}
\bibfield{author}{\bibinfo{person}{Thomas~A. Henzinger},
  \bibinfo{person}{Ranjit Jhala}, {and} \bibinfo{person}{Rupak Majumdar}.}
  \bibinfo{year}{2005}\natexlab{}.
\newblock \showarticletitle{Permissive Interfaces}.
\newblock \bibinfo{journal}{\emph{SIGSOFT Softw. Eng. Notes}}
  \bibinfo{volume}{30}, \bibinfo{number}{5} (\bibinfo{date}{sep}
  \bibinfo{year}{2005}), \bibinfo{pages}{31–40}.
\newblock
\showISSN{0163-5948}
\urldef\tempurl%
\url{https://doi.org/10.1145/1095430.1081713}
\showDOI{\tempurl}


\bibitem[\protect\citeauthoryear{Lorenzoli, Mariani, and Pezzè}{Lorenzoli
  et~al\mbox{.}}{2008}]%
        {lorenzoli18}
\bibfield{author}{\bibinfo{person}{Davide Lorenzoli}, \bibinfo{person}{Leonardo
  Mariani}, {and} \bibinfo{person}{Mauro Pezzè}.}
  \bibinfo{year}{2008}\natexlab{}.
\newblock \showarticletitle{Automatic generation of software behavioral
  models}.
\newblock \bibinfo{journal}{\emph{Proceedings - International Conference on
  Software Engineering}}, \bibinfo{pages}{501--510}.
\newblock
\urldef\tempurl%
\url{https://doi.org/10.1145/1368088.1368157}
\showDOI{\tempurl}


\bibitem[\protect\citeauthoryear{Mavridou, Rutz, and Bliudze}{Mavridou
  et~al\mbox{.}}{2017}]%
        {Mavridou19}
\bibfield{author}{\bibinfo{person}{Anastasia Mavridou},
  \bibinfo{person}{Valentin Rutz}, {and} \bibinfo{person}{Simon Bliudze}.}
  \bibinfo{year}{2017}\natexlab{}.
\newblock \showarticletitle{Coordination of Dynamic Software Components with
  JavaBIP}.
\newblock
\showISBNx{978-3-319-68033-0}
\urldef\tempurl%
\url{https://doi.org/10.1007/978-3-319-68034-7_3}
\showDOI{\tempurl}


\bibitem[\protect\citeauthoryear{Montesi}{Montesi}{2015}]%
        {montesi15}
\bibfield{author}{\bibinfo{person}{Fabrizio Montesi}.}
  \bibinfo{year}{2015}\natexlab{}.
\newblock \showarticletitle{Kickstarting Choreographic Programming}.
\newblock
\showISBNx{978-3-319-33611-4}
\urldef\tempurl%
\url{https://doi.org/10.1007/978-3-319-33612-1_1}
\showDOI{\tempurl}


\bibitem[\protect\citeauthoryear{Par\'{\i}zek and Lhot\'{a}k}{Par\'{\i}zek and
  Lhot\'{a}k}{2012}]%
        {10.1145/2384616.2384623}
\bibfield{author}{\bibinfo{person}{Pavel Par\'{\i}zek} {and}
  \bibinfo{person}{Ondřej Lhot\'{a}k}.} \bibinfo{year}{2012}\natexlab{}.
\newblock \showarticletitle{Predicate Abstraction of Java Programs with
  Collections}. In \bibinfo{booktitle}{\emph{Proceedings of the ACM
  International Conference on Object Oriented Programming Systems Languages and
  Applications}} (Tucson, Arizona, USA) \emph{(\bibinfo{series}{OOPSLA '12})}.
  \bibinfo{publisher}{Association for Computing Machinery},
  \bibinfo{address}{New York, NY, USA}, \bibinfo{pages}{75–94}.
\newblock
\showISBNx{9781450315616}
\urldef\tempurl%
\url{https://doi.org/10.1145/2384616.2384623}
\showDOI{\tempurl}


\bibitem[\protect\citeauthoryear{Raemaekers, {van Deursen}, and
  Visser}{Raemaekers et~al\mbox{.}}{2017}]%
        {RAEMAEKERS2017140}
\bibfield{author}{\bibinfo{person}{S. Raemaekers}, \bibinfo{person}{A. {van
  Deursen}}, {and} \bibinfo{person}{J. Visser}.}
  \bibinfo{year}{2017}\natexlab{}.
\newblock \showarticletitle{Semantic versioning and impact of breaking changes
  in the Maven repository}.
\newblock \bibinfo{journal}{\emph{Journal of Systems and Software}}
  \bibinfo{volume}{129} (\bibinfo{year}{2017}), \bibinfo{pages}{140--158}.
\newblock
\showISSN{0164-1212}
\urldef\tempurl%
\url{https://doi.org/10.1016/j.jss.2016.04.008}
\showDOI{\tempurl}


\bibitem[\protect\citeauthoryear{Rumbaugh, Jacobson, and Booch}{Rumbaugh
  et~al\mbox{.}}{2004}]%
        {Rumbaugh2004}
\bibfield{author}{\bibinfo{person}{James Rumbaugh}, \bibinfo{person}{Ivar
  Jacobson}, {and} \bibinfo{person}{Grady Booch}.}
  \bibinfo{year}{2004}\natexlab{}.
\newblock \bibinfo{booktitle}{\emph{Unified Modeling Language Reference Manual,
  The (2nd Edition)}}.
\newblock \bibinfo{publisher}{Pearson Higher Education}.
\newblock
\showISBNx{0321245628}


\bibitem[\protect\citeauthoryear{Voinea, Dardha, and Gay}{Voinea
  et~al\mbox{.}}{2020}]%
        {voinea20}
\bibfield{author}{\bibinfo{person}{A. Voinea}, \bibinfo{person}{Ornela Dardha},
  {and} \bibinfo{person}{Simon Gay}.} \bibinfo{year}{2020}\natexlab{}.
\newblock \bibinfo{booktitle}{\emph{Typechecking Java Protocols with
  [St]Mungo}}.
\newblock \bibinfo{pages}{208--224}.
\newblock
\showISBNx{978-3-030-50085-6}
\urldef\tempurl%
\url{https://doi.org/10.1007/978-3-030-50086-3_12}
\showDOI{\tempurl}


\end{thebibliography}

\end{document}